\PassOptionsToPackage{numbers,sort&compress}{natbib}
\documentclass[AMA, Times2COL]{WileyNJDv5} 

\usepackage{mathptmx} 
\usepackage[sfdefault]{carlito}
\usepackage{graphicx}
\usepackage{dblfloatfix}
\usepackage{setspace}
\usepackage{titlesec}
\usepackage[section]{placeins}

\articletype{{\bfseries Research article}}%

\received{Date Month Year}
\revised{Date Month Year}
\accepted{Date Month Year}
\journal{Laser \& Photonics Reviews}
\volume{00}
\copyyear{2025}
\startpage{1}

\newcommand{\printkeywords}{
	\bmsection*{Keywords}
	Optical computing, nonlinear operations, kernel regression, reservoir computing
}

\newcommand{\myabstracttext}{
	Optical kernel machines offer high throughput and low latency. A nonlinear optical kernel can handle complex nonlinear data, but power consumption is typically high with the conventional nonlinear optical approach. To overcome this issue, we present an optical kernel with structural nonlinearity that can be continuously tuned at low power. It is implemented in a linear optical scattering cavity with a reconfigurable micro-mirror array. By tuning the degree of nonlinearity with multiple scattering, we vary the kernel sensitivity and information capacity. We further optimize the kernel nonlinearity to best approximate the parity functions from first order to fifth order for binary inputs. Our scheme offers potential applicability across photonic platforms, providing programmable kernels with high performance and low power consumption.
}

\titleformat{\section}
{\large\sffamily\bfseries} 
{\thesection.} 
{0.5em} 
{} 

\titleformat{\subsection}
{\normalfont\sffamily\normalsize\bfseries} 
{\thesubsection.}
{0.5em}
{}

\titleformat{\subsubsection}
{\normalfont\sffamily\normalsize\itshape} 
{\thesubsubsection.}
{0.5em}
{}

\begin{document}
	
	\title{{\fontsize{24}{28}\selectfont\bfseries\sffamily
			Optical kernel machine with programmable nonlinearity}}
	
	\authormark{}
	\titlemark{{RESEARCH ARTICLE}}
	
	
	\abstract{{
			\myabstracttext
	}}
	
	\jnlcitation{\cname{%
			\author{SeungYun Han},
			\author{Fei Xia},
			\author{Sylvain Gigan},
            \author{Bruno Loureiro},
			\author{Hui Cao*}}.
		\ctitle{Tunable Nonlinear Optical Mapping in a Linear Scattering Cavity.} \cjournal{
		} \cvol{2025;XX(XX):X--X}.}
	
	\newcommand{\customauthors}{
		{\Large\itshape\mdseries\sffamily 
			SeungYun Han, 
			Fei Xia, 
			Sylvain Gigan, 
            Bruno Loureiro, and 
			Hui Cao*}
	}
	
	\newcommand{\fullauthorfootnote}{%
		\parbox{\linewidth}{
			\fontsize{9.5pt}{11.4pt} 
			\raggedright 
			\sffamily
			SY.Han, H.Cao \\
			Applied Physics Department \\
			Yale University \\
			Connecticut, USA\\
			E-mail: seungyun.han@yale.edu; hui.cao@yale.edu
            \vspace{4pt}
			\par
			F. Xia, S. Gigan \\
            Laboratoire Kastler Brossel \\
            ENS-Universite PSL, CNRS \\
            Sorbornne Universit\'e, Coll\`ege de France \\
            Paris, France
            \vspace{4pt}
			\par
			F. Xia \\
            Department of Electrical Engineering and Computer Science \\
            University of California, Irvine \\
            California, USA
            \vspace{4pt}
			\par
            B. Loureiro \\
            D\'epartement d'Informatique \\
            \'Ecole Normale Sup\'erieure - PSL \& CNRS \\
            Paris, France
            \vspace{4pt}
			\par
			\vspace{0.5em} 
			*Corresponding author: \texttt{hui.cao@yale.edu}%
		}%
	}

	\renewcommand{\thefootnote}{\fnsymbol{footnote}}
	\makeatletter
	\renewcommand{\maketitle}{
		\twocolumn[
		{
			\vspace*{1.5em}
			\noindent\@title\par
			\vspace{0.8em}
			\customauthors\par
			\vspace{1.2em}
			
			\noindent
			\begin{minipage}[t]{0.67\textwidth}
				\colorbox{gray!42}{
					\begin{minipage}{0.95\linewidth}
						\sffamily
						\normalsize
						\begin{spacing}{1.1}
							\vspace*{0.5em}
							\myabstracttext
							\vspace*{-1em}
						\end{spacing}
					\end{minipage}
				}            
			\end{minipage}
			\hspace{-0.3em} 
			\begin{minipage}[t]{0.3\textwidth}
			\end{minipage}
			\bigskip
		}
		]
	}
	\makeatother
	
	\renewcommand{\thefootnote}{\fnsymbol{footnote}}
	
	\maketitle
	
	\footnotetext[1]{\fullauthorfootnote}
	
	\section{Introduction}\label{introduction}
	
	\FloatBarrier
	\begin{figure*}[!b]
		\centerline{\includegraphics[width=0.95\textwidth,
			trim=120pt 25pt 18pt 52pt,clip]{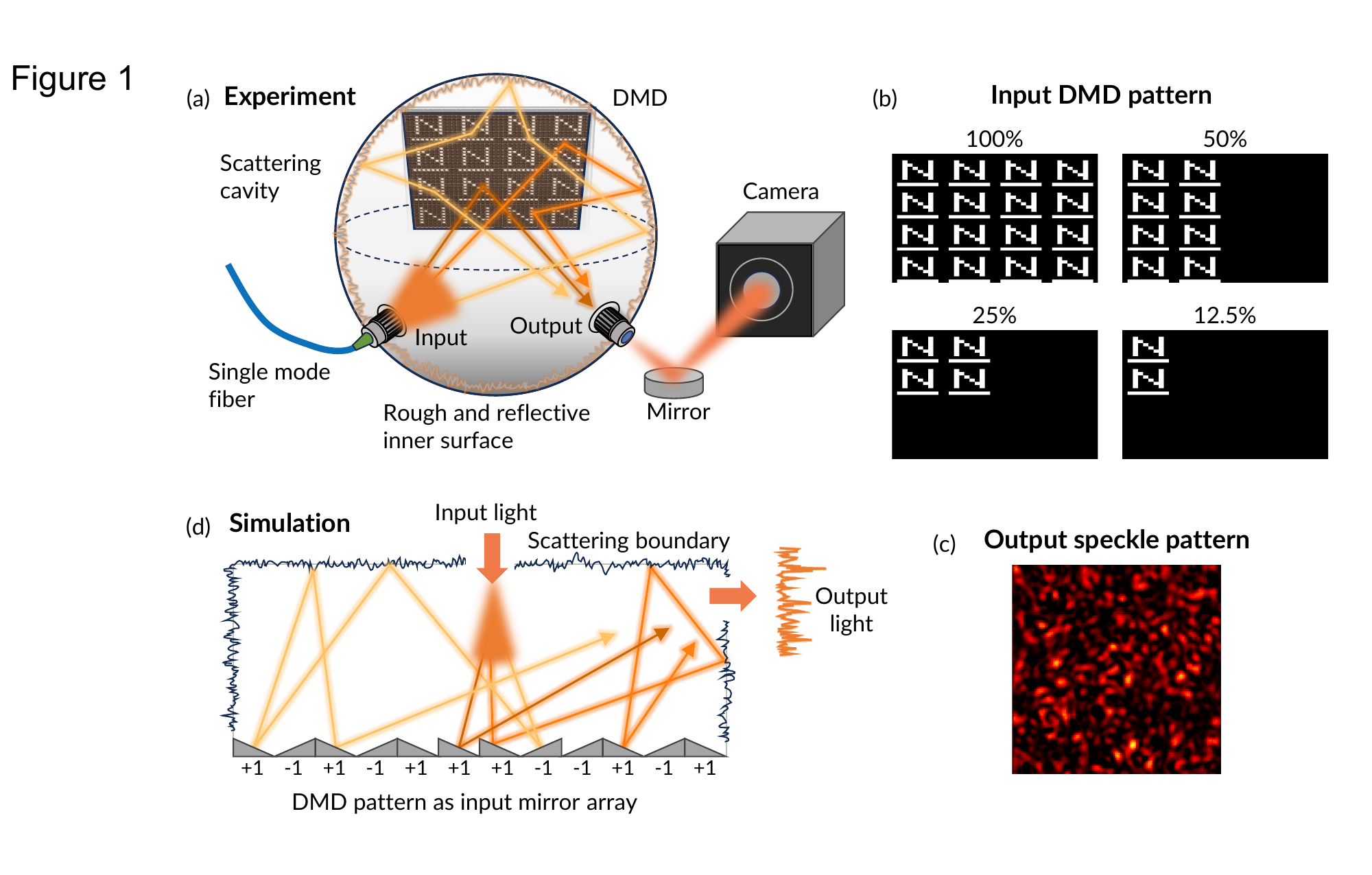}} 
		\caption{\fontsize{9.5}{11.5}\selectfont Programmable nonlinear optical kernel. 
			(a) Schematic of experimental implementation of the kernel in a linear scattering cavity. Input light at $\lambda$ = 1550 nm is scattered by the rough cavity wall and a reconfigurable micro-mirror array (DMD). Part of output light is recorded by a CCD camera. 
            (b) Four binary masks written to the DMD. The micro-mirrors (pixels) tilt at angles of $+12^\circ$ and $-12^\circ$, redirecting incident light in 2 directions, each representing 1 and -1 in the input pattern. A representative base pattern with $10 \times 10$ macropixels is repeated to occupy $100\%, 50\%, 25\%, 12.5\%$ area of the DMD. 
			(c) Measured speckle intensity pattern of output light from the scattering cavity.
			(d) Schematic of numerically-simulated 2D scattering cavity with varying reflectivity of the wall. The cavity dimension is $130 \lambda \times 60 \lambda$. One side of the cavity wall is covered by 12 mirrors, each can be flipped to $+15^\circ$ or $+15^\circ$ angle. 
			 \label{fig_schematic}}
	\end{figure*}
	
	Kernel machine learning is a powerful approach to learning complex, nonlinear patterns by mapping data to higher-dimensional spaces~\cite{aronszajn1950theory, cortes1995support, scholkopf2002learning, shawe2004kernel, rahimi2007random}. It is highly effective with limited data, offering strong performance, robustness, and good generalization~\cite{arora2019harnessing, brigato2021close, mallick2021deep}. Kernel methods are especially advantageous when the relationships in the data are not linearly separable. However, they often scale poorly with large datasets because kernel matrices grow quadratically with the number of samples, leading to high memory and computation costs~\cite{scholkopf2002learning, rahimi2007random, le2014fastfood}.
	
	Optical implementation of kernel machine learning involves the realization of kernel-based computations directly with light. These implementations can achieve extremely high-speed parallel processing with significantly reduced energy consumption~\cite{brunner2013parallel, rafayelyan2020large, sunada2020using, pierangeli2021photonic, zhou2022nonlinear, hu2023tackling, ccarpinliouglu2025genetically, desnos2025phase}. This makes them ideal for edge computing scenarios where latency and power efficiency are critical, such as high-speed signal classification~\cite{vandoorne2014experimental, saade2016random, ohana2020kernel, pierangeli2021photonic, cong2024implementing, foradori2025neuromorphic, zhou2025deep}, chaotic system prediction~\cite{larger2012photonic, paudel2020classification, rafayelyan2020large, sunada2020using}, signal recovery in optical communication~\cite{argyris2018photonic, sozos2022high, silva2025integrated}, and ultrafast image or spectrum recognition~\cite{denis2022photonic, zhou2023ultrafast}. However, free-space optical systems often suffer from alignment sensitivity, and on-chip ones from fabrication imperfections and limited reconfigurability. 
	
	To build a feature-space embedding best suited for a specific task, we explore an optical kernel machine with tunable nonlinearity. In particular, we utilize the structural nonlinearity in a linear multiple-scattering cavity because it allows tuning the nonlinear order without the need of changing optical power~\cite{eliezer2023tunable, xia2024nonlinear}. By encoding the input to the scattering structure rather than the illumination light, the output light of the cavity has a nonlinear relation with the input even without material or detection nonlinearities. The order of structural nonlinearity can be tuned continuously with cavity parameters at constant low power. 
	
	We present two methods of turning the structural nonlinearity and show that stronger nonlinearity leads to higher sensitivity of the kernel output to the input change. This allows us to tune the kernel sensitivity for the learning task. High sensitivity leads to high variance and low bias; thus, the kernel is very flexible and can capture intricate nonlinear patterns in the data. A less sensitive kernel, on the other hand, is more rigid and robust to noise. We find that the information capacity of our optical kernel grows with structural nonlinearity. Information capacity is a measure of the expressive power of the kernel that reflects the complexity of the functions it can represent. By tuning the structural nonlinearity, we show that the nonlinear optical kernel can best approximate target nonlinear functions of varying complexity, e.g., parity functions of different orders for binary inputs.

	\section{Tunable optical kernel}\label{Results and Discussion}

	\subsection{Multiple-scattering cavity}\label{sec_setup_design}
    
	There are different ways of realizing structural nonlinearity, which have been used for optical implementation of artificial neural networks~\cite{faqiri2022physfad, momeni2023backpropagation, xia2024nonlinear, yildirim2024nonlinear, wanjura2024fully, li2024nonlinear, abou2025programmable, richardson2025nonlinear, you2025nonlinear, liu2025incoherent}. We use a multiple-scattering cavity to achieve a nonlinear mapping between the scattering potential and the output light \cite{eliezer2023tunable}. 
	
	Our experimental setup is shown schematically in Figure~\ref{fig_schematic}a. A continuous-wave linearly-polarized light from a single-frequency laser (Keysight Technologies 81940A) at wavelength $\lambda = 1550$ nm and power 21.3 mW injected into a spherical cavity of diameter 3.75 cm through a single-mode fiber. The opposite side of the cavity wall has an open window covered by a digital micro-mirror device (DMD). The DMD (Texas Instruments DLP9000X) modulation area within the window has 1280 $\times$ 800 micro-mirrors, each can be flipped to either $+12^{\circ}$ or $-12^{\circ}$. The inner surface of the cavity is rough and highly reflective. The input light is scattered multiple times by the cavity wall and the DMD. Part of the light escapes from the cavity through an opening and is directed by a mirror to an InGaAs camera (Xenics Xeva FPA-640). A linear polarizer is placed in front of the camera that records the output intensity pattern. 
	
	At low power, the output field $E_{\rm out}$ scales linearly with the input field $E_{\rm in}$ and can be expressed in the Born series:
	\begin{equation}\label{Born series}
	E_{\rm out} = \left[ V + V(G_0 V) + V(G_0 V)^2 + ...\right]E_{\rm in} \, ,
	\end{equation}
	\noindent where $V$ represents optical scattering potential inside the cavity, and $G_0$ describes free-space propagation between subsequent bounces from cavity wall or DMD. The different orders of $V$ in Eq.\ref{Born series} describe the number of bounces light experiences before leaving the cavity. $V$ is the sum of scattering potential of the DMD $V_d$ and that of the cavity wall $V_c$, $V= V_d + V_c$. The DMD is reconfigurable, each configuration producing a unique output speckle pattern (Figure~\ref{fig_schematic}c). With multiple bounces from the DMD, the mapping from $V_d$ to $E_{\rm out}$ is nonlinear. We term such nonlinearity as structural nonlinearity.
	
	There are two ways of varying the degree of structural nonlinearity. One is changing the number of bounces off the DMD. It can be realized by varying the lifetime of light inside the cavity with tunable wall reflectivity. Since this is difficult to realize experimentally, we numerically simulate a two-dimensional (2D) cavity with adjustable wall reflectivity (Figure~\ref{fig_schematic}d) and show in the following section the tuning of structural nonlinearity. Another way of tuning structural nonlinearity is varying the effective modulation area on the DMD. This will change the relative weight of the reconfigurable scattering potential to the fixed one (Figure~\ref{fig_schematic}b). A larger DMD modulation area leads to a higher probability that light hits it in each bounce, resulting in a stronger structural nonlinearity. This method is adopted in our experiment. 
	
	\subsection{Structural nonlinearity} \label{sec_linear_regression}

	\begin{figure}[!t]
		\centerline{\includegraphics[width=0.425\textwidth,
			trim=245pt 15pt 398pt 32pt,clip]{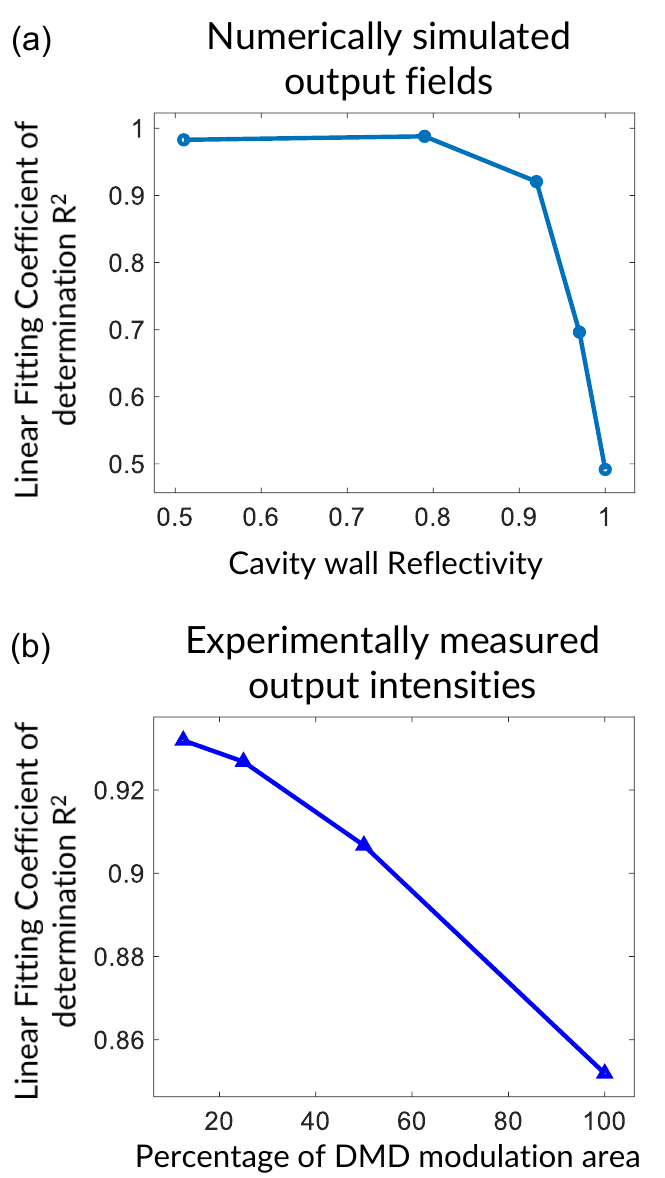}} 
		\caption{\fontsize{9.8}{11.8}\selectfont Tunable structural nonlinearity in a linear multiple-scattering cavity. Linear regression of cavity input-output mapping gives the coefficient of determination $R^2$. (a) In the numerically-simulated 2D cavity $R^2$ decreases with increasing cavity wall reflectivity. (b) In the experimental 3D cavity, $R^2$ drops as the DMD modulated area increases. The decrease of $R^2$ is a result of enhanced structural nonlinearity. }\label{fig_linearFit}
	\end{figure}
    
	The simplest way to characterize the strength of structural nonlinearity is to quantify its deviation from linear regression. In our numerical simulation of the 2D scattering cavity, the wall reflectivity is set to $r=0.51, 0.78, 0.92, 0.97, 1$. For binary inputs ($\pm 1$) to 12 switchable mirrors ($\pm 15^\circ$) in the cavity, the total number of mirror configurations is $K=2^{12}$. For each configuration, we calculate the one-dimensional output field pattern $E_{\rm out}(r)$ that contains 42 speckle grains. Then we fit a linear model $\hat{E}_{\rm out} = W \, x$, where $x$ represents the binary input, and quantify the deviation with the coefficient of determination 
    \begin{align}
        R^2 \;=\; 1 \;-\; \frac{\mathrm{Var}(E_{\rm out} - \hat{E}_{\rm out})}{\mathrm{Var}(E_{\rm out})}.
    \end{align}   
    Figure~\ref{fig_linearFit}a shows that $R^2$ decreases continuously with increasing reflectivity $r$ of the cavity wall. A higher reflectivity extends the lifetime of light inside the cavity, allowing more bounces of light off the mirror array before leaving the cavity, thus the output fields contain more nonlinear orders. 
    
	Experimentally, we vary the DMD modulation area. A binary pattern of dimension $M$ is repeatedly written spatially to occupy a certain percentage of the total area of the DMD. Outside this area, the micro-mirrors are set to -1 ($-12^\circ$). When the DMD modulation area increases, the base pattern is fixed in size but is repeated additional times. In Fig.~\ref{fig_schematic}c, the base pattern contains $M=10\times10$ macropixels, each comprising $16 \times 10$ pixels (micro-mirrors). All pixels within a macropixel are set to the same value of 1 ($+12^\circ $) or -1 ($-12^\circ $). The DMD modulation area is set to $12.5\%, 25\%, 50\%, 100\%$. Since the total number of DMD configurations is too large ($2^{100}$), we randomly sample $K=2000$ patterns. The output intensity pattern recorded by the camera contains 258 speckle grains. 
	
	Since part of the input light is not scattered by the DMD modulated area at all, its output forms a constant background. This background field $E_{\rm bg}$ contributes to a linear response of the output intensity to the DMD pattern. We denote the output field that varies with the DMD configuration as $E_{mod}$, and the total output field $E_{\rm out} = E_{\rm bg} + E_{\rm mod}$. Thus, the output intensity is
	\begin{equation}
	I_{\rm out} = |E_{\rm bg}|^2 + |E_{\rm mod}|^2 + 2 \text{Re}[E_{\rm bg}^*E_{\rm mod}].
	\end{equation}
	The interference between the background and modulated fields leads to the cross term that scales linearly with $E_{\rm mod}$. The background intensity $I_{\rm bg} = |E_{\rm bg}|^2 + \langle|E_{\rm mod}|^2\rangle_K$ is obtained by averaging the output intensity over random inputs to the DMD, and is subtracted from $I_{\rm out}$. Since $E_{\rm mod}$ contains the single scattering term that scales linearly with $V_d$ (Eq.~\ref{Born series}), $I_{\rm out}-I_{\rm bg}$ has a linear term of $V_d$.     

	In Fig.~\ref{fig_linearFit}b, as the DMD modulation area increases, $I_{\rm out}-I_{\rm bg}$ shows a stronger deviation from linear regression with DMD inputs. This reflects a stronger structural nonlinearity with a larger DMD modulation area. 
	
	A more quantitative assessment of the structural nonlinearity can be done with Boolean function analysis~\cite{o2014analysis}. For binary inputs $x$ of dimension $M$, any output $y$ can be decomposed into terms of different orders:
	\begin{align}
		\varphi &= f(x_1,\cdots,x_M) \notag\\
		&= c_0^{(0)} + \sum_{m_1=1}^{M}c_{m_1}^{(1)} \, x_{m_1} + \sum_{m_1=2}^{M}\sum_{m_2=1}^{m_1-1}c_{m_1, m_2}^{(2)} \, x_{m_1}x_{m_2} \notag\\
		& \ \ \ \ \ + \cdots+ c_{1,2,\cdots,M}^{(M)} \, x_{1}x_2\cdots x_{M},
	\end{align}
	where $x_m$ is the $m$-th element of the input, and $c^{(n)}_{m_1, \cdots, m_n}$ is the $n$-th order coefficient of the term $x_{m_1}\cdots x_{m_n}$. The magnitude of the coefficient represents the contribution of the corresponding term to the output. Therefore, to evaluate the contributions of different orders, we summed the absolute value of the coefficients for each $n$:
	\begin{align}
		S_n = \sum_{m} \left|\tilde{c}^{(n)}_m\right|
	\end{align}
	where $m$ abbreviates $m_1,\cdots,m_n$, and $\tilde{c}^{(n)}_m$ is the normalized coefficient
	\begin{align}
		\tilde{c}^{(n)}_m = \frac{c^{(n)}_m}{\sum_{n}\sum_{m} \left|c^{(n)}_m\right|}.
	\end{align}   
	The results of the Boolean function analysis will be presented in the later section of kernel regression. 
	
	\subsection{Kernel expressivity} \label{sec_correlation_transfer}
    
	Our multiple-scattering cavity performs a nonlinear random projection that embeds the dataset to a high-dimensional feature space, i.e., a \emph{random feature map}. Here, we evaluate how sensitive the output speckle pattern is to the input change. The change in input or output is quantified by the normalized correlation function:
	\begin{align}
		\text{C}(z_1 ,z_2) = \frac{ z_1^\dagger \cdot z_2}{(z_1^\dagger \cdot z_1)^{1/2} \, (z_2^\dagger \cdot z_2)^{1/2}},
	\end{align}
	where $z_1, z_2$ represent a pair of input or output vectors. 
	
\begin{figure}[!]
		\centerline{\includegraphics[width=0.45\textwidth,
			trim=200pt 35pt 428pt 12pt,clip]{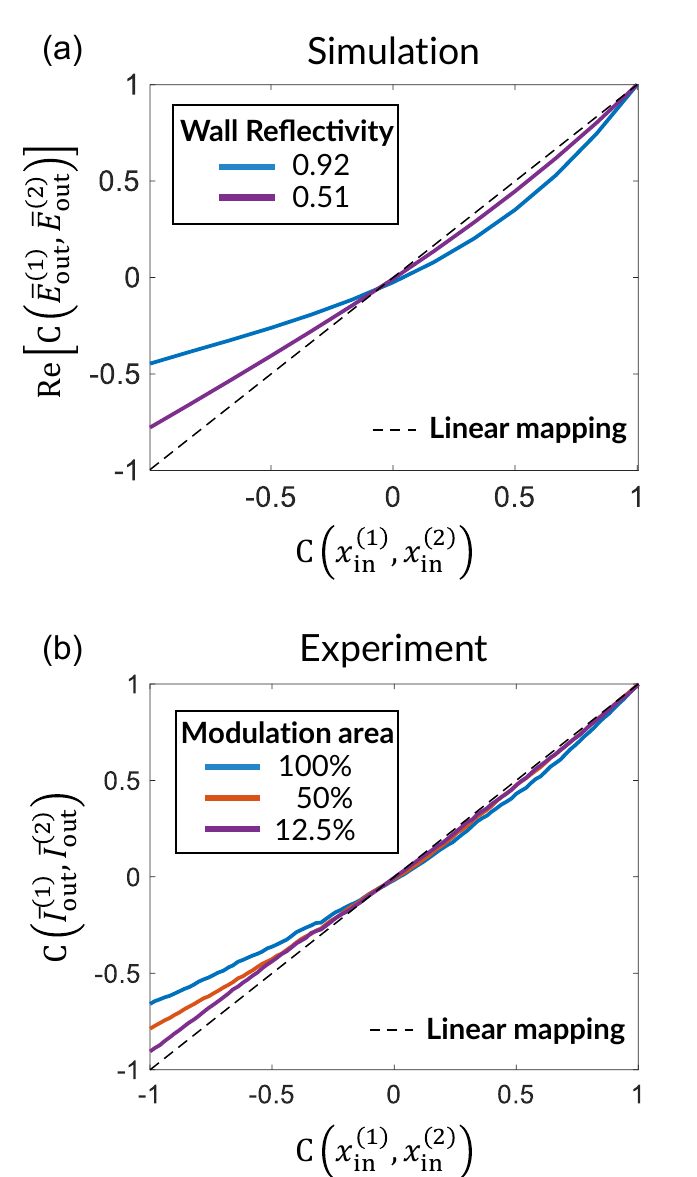}} 
		\caption{\fontsize{9.8}{11.8}\selectfont Output sensitivity to input change. (a) Real part of the correlation between output field patterns for different mirror configurations increases with the correlation between input patterns (mirror configurations) in the simulated 2D cavity with wall reflectivity of 0.92 (blue) and 0.51 (purple). The black dashed curve is for linear input-output mapping. For higher wall reflectivity, the curve deviates more from the linear mapping due to stronger structural nonlinearity.  (b) Correlation of output intensity patterns from the experimental 3D cavity increases with the correlation of input patterns on the DMD with the modulation area of 100\% (blue), 50\% (orange), and 12.5\% (purple). More deviations from the linear mapping (black dashed line) result from stronger structural nonlinearity at larger DMD modulation area. }
		\label{fig_correlation_InOut}
	\end{figure}

	For the simulated 2D scattering cavity, the input pattern on the 12 mirrors is a binary vector of dimension 12. The output field pattern $E_{\rm out}(r)$, which has 42 speckle grains, is sampled at 549 positions, producing a complex vector of dimension 549. Then the mean output field from $K=2^{12}$ samples $\langle E_{\rm out} \rangle_K$ is subtracted from the output field $E_{\rm out}$ to obtain the feature map (output vectors) $\varphi = \bar{E}_{\rm out} =E_{\rm out}-\langle E_{\rm out} \rangle_K$. The cross-correlation of two output vectors corresponding to different mirror configurations (input vectors) is, in general, a complex number, but its imaginary part vanishes when the number of speckle grains is very large. This is because mathematically $C^*_{\rm out}(\varphi_1, \varphi_2) = C_{\rm out}(\varphi_2, \varphi_1)$ and statistically $C_{\rm out}(\varphi_2, \varphi_1) = C_{\rm out}(\varphi_1, \varphi_2)$ for random output fields. In Fig.~\ref{fig_correlation_InOut}a, we plot the real part of the output field correlation ${\rm Re}[C_{\rm out}]$ versus the input correlation $C_{\rm in}$, for weak and strong structural nonlinearity (corresponding to low and high cavity wall reflectivity). Note that a linear feature map would imply a linear relationship between $C_{\rm in}$ and ${\rm Re}[C_{\rm out}]$ (black dashed line). Both curves start at \((1,1)\) since identical inputs produce identical outputs. As the input correlation decreases, the output correlation falls more rapidly with stronger structural nonlinearity. The accelerated output decorrelation indicates higher output sensitivity to input change, namely, the kernel nonlinearity is enhanced by structural nonlinearity. Both curves pass through $(0,0)$, which means that uncorrelated inputs generate uncorrelated outputs. As the input correlation becomes negative, the output correlation also goes negative, but with a smaller magnitude. This indicates that the nonlinear input-output mapping contains polynomial terms of even order, i.e., $x^n$ with even $n$. With stronger nonlinearity, there are more terms with even order, and the output correlation further deviates from -1 when the input is anti-correlated ($C_{\rm in} = -1 $). 
	
	Next, we switch to the experimental three-dimensional (3D) scattering cavity. For a base pattern of dimension $M$ on the DMD, the input binary vector is of dimension $M$. The 2D output intensity pattern is converted to an 1D vector of dimension $N$. The mean intensity of all outputs from $K$ samples $\langle I_{\rm out} \rangle_K$ is subtracted from the output intensity $I_{\rm out}$, and the resulting zero-mean intensity is used as the feature map $\varphi=\bar{I}_{\rm out} =I_{\rm out}-\langle I_{\rm out} \rangle_K$. Figure~\ref{fig_correlation_InOut}b shows the correlation of the output intensity $C_{\rm out}$ as a function of the input correlation $C_{\rm in}$, for different DMD modulation areas. With a larger modulation area, the stronger structural nonlinearity makes the curve deviate more from the straight line for linear input-output mapping. The output's reduced correlation or memory of the input allows a more diverse response of the kernel machine. 
	
	The high sensitivity of output to input change reflects the high expressivity of a kernel machine, which leads to large variance and low bias. While high variance means that the model is very flexible and can capture intricate, nonlinear patterns in the data, it is also highly sensitive to minor fluctuations and noise in the training set. In contrast, low sensitivity leads to high bias and low variance. A less sensitive model is more rigid and makes stronger assumptions about the data, which may cause it to miss important patterns. 
	
	Our kernel sensitivity can be tuned and optimized for a specific task, depending on the task's complexity and the noise level of the dataset. This will make the kernel machine accurate for the task without overfitting. 
	
	\subsection{Information capacity} \label{sec_info_capacity}
	
	In this section, we consider the information capacity of the nonlinear kernel. Information capacity \cite{kuang2025bounds} characterizes the number of independent, noise-resolvable output modes of the kernel~\cite{cox1986information, miller2000communicating}. It informs the performance and complexity trade-offs of kernel machines, primarily through the Information Bottleneck principle\cite{tishby2000information}. By viewing a kernel machine as a communication channel, the concept of capacity helps regulate the amount of relevant information preserved by the kernel, thereby influencing generalization and complexity\cite{hu2022quantifying}. 
    
	Information capacity is defined from the singular value spectrum of responses $\varphi$ (cavity outputs) under a constant signal-to-noise ratio (SNR). Specifically, the responses (output fields or intensities) $\{\varphi_k\} $ to $K$ random binary inputs are assembled into a response matrix 
	\begin{align}
         \boldsymbol{\Phi} = [\,\varphi_1,\;\varphi_2,\;\dots,\;\varphi_K\,].
	\end{align}
	The singular value decomposition of the response matrix gives:
	\begin{align}
	     \boldsymbol{\Phi} = \mathbf{U}\,\boldsymbol{\Sigma}\,\mathbf{V}^{T},	\qquad
        \boldsymbol{\Sigma} = \mathrm{diag}(\sigma_1,\sigma_2,\dots,\sigma_s).
	\end{align}
	where $s$ is the number of singular values. We normalize the singular values
	\begin{align}
		\tilde{\sigma}_i = 
		\frac{\sigma_i}{\sqrt{\sum_{j=1}^s \sigma_j^2}}.
	\end{align}
	Consider additive white noise and $Q$ the signal-to-noise ratio (SNR), the information capacity is given by~\cite{kuang2025bounds}
	\begin{align}
		P =  \sum_{i=1}^s \log_{2}\!\bigl(1 + Q \, \tilde{\sigma}_i^{\,2}\bigr).  
	\end{align}
	This quantity is the effective number of channels of the kernel above the noise level, which is closely related to the expressive capacity of physical learning machines\cite{hu2022quantifying}. This is also closely connected to the \emph{degrees of freedom} of the kernel, a quantity often appearing in the statistical analysis of kernel ridge regression and providing a notion of the effective dimension of the kernel feature space\cite{zhang2005learning,caponnetto2007optimal}.
	
\begin{figure}[h]
		\centerline{\includegraphics[width=0.4\textwidth,
			trim=200pt 45pt 468pt 32pt,clip]{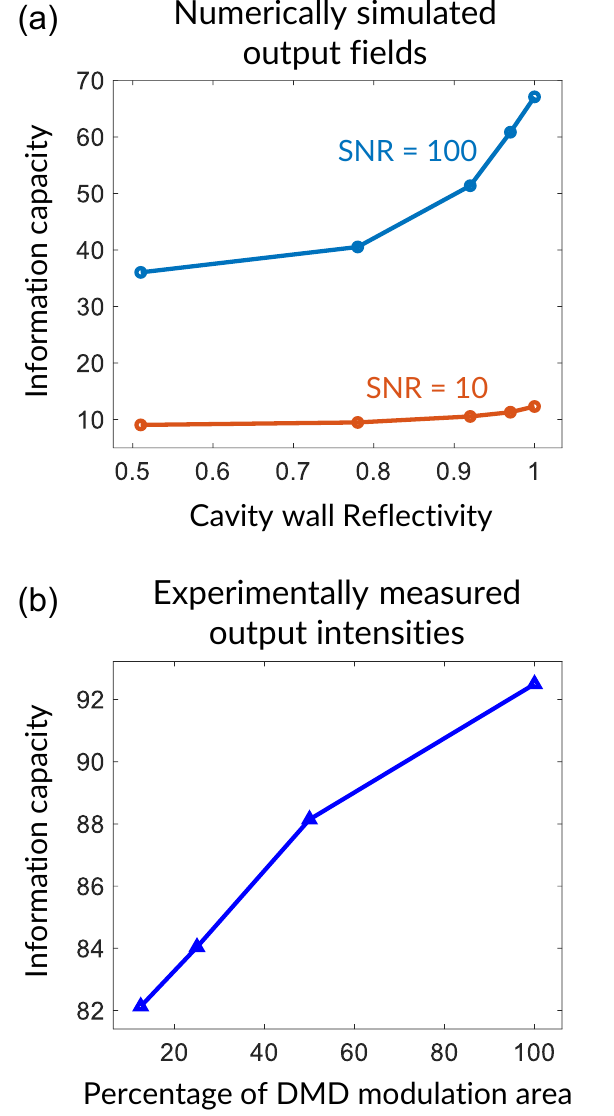}} 
		\caption{\fontsize{9.8}{11.8}\selectfont Information capacity of the multiple-scattering cavity. (a)
			Information capacity increases with the wall reflectivity of the simulated 2D cavity. SNR = 100 (blue) leads to a larger information capacity than SNR = 10 (orange). (b) Information capacity increases with the DMD modulation area in the experimental 3D cavity with $\mathrm{SNR}=95.2$.}
		\label{fig_informationCapacity}
	\end{figure}

	The information capacity $P$ for the simulated 2D scattering cavity is shown in Fig.~\ref{fig_informationCapacity}a as a function of the cavity wall reflectivity $r$ for $Q$ = 10, 100. $P$ increases with $r$, indicating that the structural nonlinearity increases the information capacity. Higher SNR leads to larger information capacity, as expected.
	Figure~\ref{fig_informationCapacity}b shows that the information capacity $P$ of the experimental 3D scattering cavity increases with the DMD modulation area. Again, a higher nonlinearity yields a larger information capacity, as more singular modes become significant above the noise level. These results demonstrate that the tuning of the structural nonlinearity controls the channel capacity and degrees of freedom of the kernel.
                
	\section{Kernel Regression} \label{sec_kernel_regression}

	\begin{figure*}[b]
		\centerline{\includegraphics[width=0.92\textwidth,
			trim=200pt 145pt 245pt 110pt,clip]{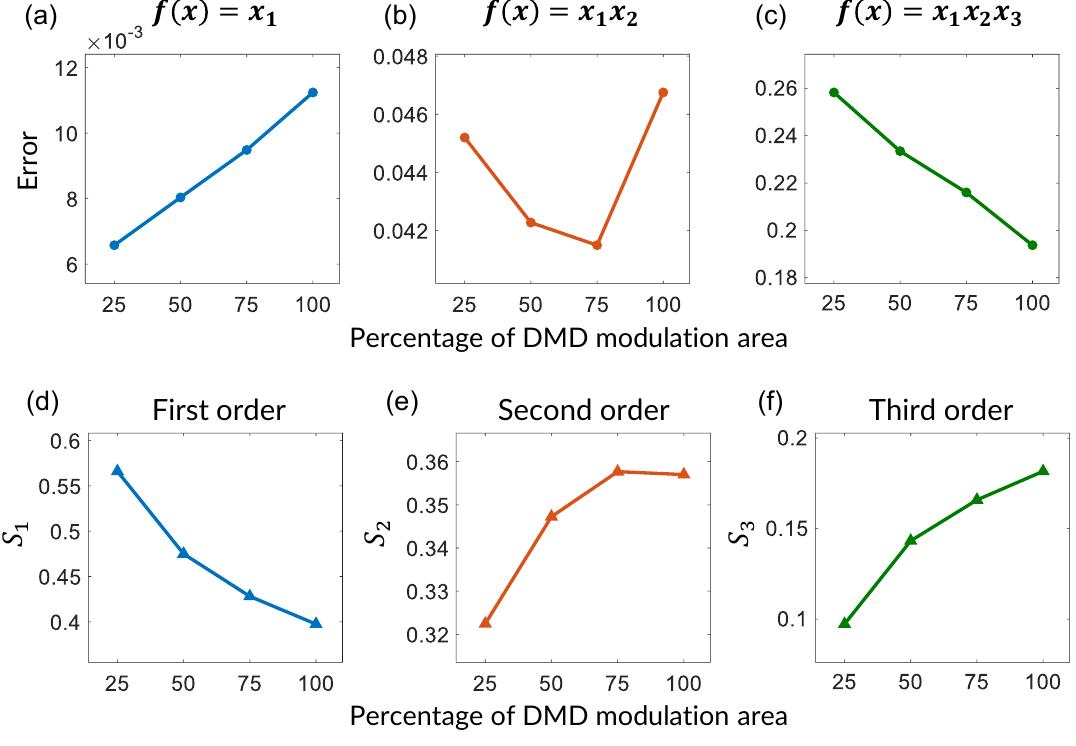}}  
		\caption{\fontsize{9.8}{11.8}\selectfont Kernel regression to parity functions of varying order. The experimental 3D cavity serves a kernel by providing a high-dimensional feature map that maps input patterns on the DMD to optical intensity patterns at cavity output. Binary patterns $(x_1, x_2, ..., x_9)$ are uploaded to the DMD, and repeated multiple times to occupy a certain percentage of the entire modulation area. Two-dimensional output intensity pattern is linearly regressed to parity function functions of first order \( f(x) = x_1 \) in (a), second-order \( f(x) = x_1 x_2 \) in (b), and third-order \( f(x) = x_1 x_2 x_3 \) in (c). The regression error is plotted as a function of the percentage of DMD modulation area, which controls the kernel nonlinearity. The measured intensity pattern consists of approximately 55 speckle grains. (d-f) Magnitude sum of Boolean function coefficients of first order $S_1$ in (d), second-order $S_2$ in (e), and third-order $S_3$ in (f) are computed for varying percentage of DMD modulation area. The regression error for the parity function of order $i$ reaches the minimum at the DMD modulation area where $S_i$ is maximum. Kernel regression has the best performance when its nonlinearity, tuned by the DMD modulation area, matches the order of the parity function. 
		}
		\label{fig_kernelRegression}
	\end{figure*}
    
	The key idea of kernel regression is to map inputs to a high-dimensional feature space, where the target can be extracted with a linear model. We employ our multiple-scattering cavity for the physical mapping of the inputs on the DMD to the feature space of output speckles, replacing the digital computation with optical measurements. Then, a linear regression from the feature space (speckle pattern) is performed to estimate the target function. Since only a linear model is applied in the digital domain, the nonlinear capacity of the kernel regression originates from the physical kernel. Instead of using materials with nonlinear optical responses \cite{van2017advances, ryou2021free, teugin2021scalable, wright2022deep, li2023all, mcmahon2023physics, wang2023image, nikkhah2023reconfigurable, wang2024large, momeni2025training, saeed2025nonlinear, yanagimoto2025programmable, wu2025field} or optical-electronic conversion \cite{hughes2018training, wang2022optical}, the programmable nonlinearity in our cavity allows changing the characteristics of the feature space mapping, providing a distinct advantage of optimizing the kernel for different target functions.
	
	As an example, we consider the parity function of varying order, also known as \emph{sparse parity}, as the target function. For binary variables $x_i = -1$ or $1$, $i = 1, 2, ....M$, the $l$-th order or $l$-sparse parity function is $ f(x)= \prod_{i=1}^{l} x_i$. This function is highly variable and is notoriously difficult to learn with a kernel method, especially when the order $l$ is equal to the input dimension $M$~\cite{blum2003noise,bengio2005curse}. Therefore, it provides an ideal testbed to study the degree of nonlinearity implemented by our optical kernel. 
    
	We start with the input dimension $M=9$ by uploading $3\times3$ binary patterns to the DMD. We record the output speckle patterns of the 3D scattering cavity for all possible input patterns ($K=2^{9}$). During measurement, the order of input patterns is randomized to minimize any influence of temporal drift of the output pattern. Each output pattern contains 36 speckle grains. We randomly choose 90\% of the data for training of the linear regression and the remaining 10\% for testing. Performance is quantified by the root-mean-square error (RMSE) on the test set:
	\begin{align}
		\mathrm{RMSE}_\mathrm{test}
		= \sqrt{\frac{1}{N_{\mathrm{test}}} 
			\sum_{j=1}^{N_{\mathrm{test}}} 
			\bigl(y_j - \hat{f}(x_j)\bigr)^{2}},
	\end{align}
	\noindent where $N_{\mathrm{test}}$ is the number of test patterns, $\hat{f}(x_j)$ is the prediction, and $y_j$ is the ground truth. 
	
	To minimize the test error for a parity function of a specific order, we optimize the kernel nonlinearity by varying the DMD modulation area. For the first-order parity function $f(x) = x_1$, the test error $\mathrm{RMSE}_\mathrm{test}$ is minimum at the smallest (25\%) DMD modulation area (Fig.~\ref{fig_kernelRegression}a). This is because a smaller area generates the more linear input-output relation, which better matches the linear target function. This is further confirmed in the Boolean function analysis. As shown in Fig.~\ref{fig_kernelRegression}d, the first-order contribution $S_1$ to kernel output is larger at a smaller DMD modulation area. When the target function becomes nonlinear, e.g., the second-order parity function $f(x) = x_1 \, x_2$, the test error $\mathrm{RMSE}_\mathrm{test}$ first decreases and then increases with the DMD modulation area (Fig.~\ref{fig_kernelRegression}b). The Boolean function analysis reveals that the second-order contribution $S_2$ to the kernel output first increases and then slightly decreases with the modulation area (Fig.~\ref{fig_kernelRegression}e). With 75\% of the DMD area being modulated, $S_2$ reaches the maximum, and the kernel best approximates the second-order parity function. For a higher-order nonlinear function, e.g., the third-order parity function $f(x) = x_1 \, x_2 \, x_3$, the test error $\mathrm{RMSE}_\mathrm{test}$ decreases monotonically with increasing modulation area  (Fig.~\ref{fig_kernelRegression}c). According to the Boolean function analysis, the third-order contribution $S_3$ to the kernel output increases monotonically with the modulation area (Fig.~\ref{fig_kernelRegression}f). Hence, stronger structural nonlinearity can better approximate higher-order parity functions.  
	
	In the above analysis, the input and output dimensions of the kernel are fixed when varying the DMD modulation area. Next, we set the DMD modulation area to 100\%, and increase the dimension of input patterns on the DMD to approximate higher-order nonlinearity functions. We simultaneously vary the number of output speckles detected by the camera to ensure that the output dimension is not the limiting factor for the kernel performance. 
    
	\begin{figure}[!h]
		\centerline{\includegraphics[width=0.42\textwidth,
			trim=210pt 55pt 438pt 22pt,clip]{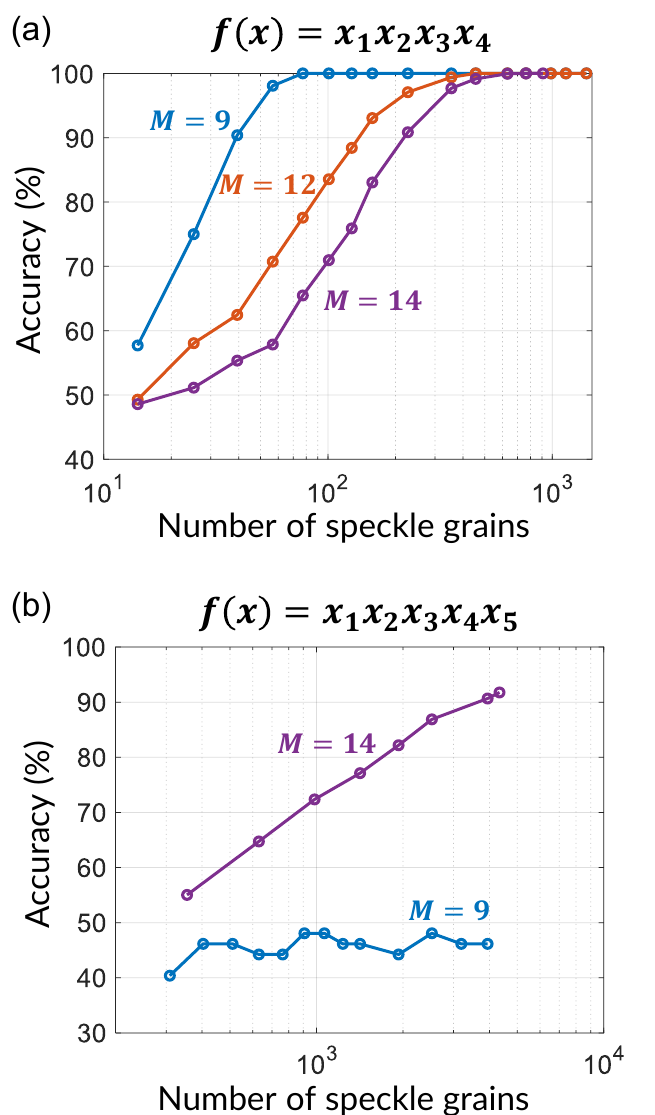}} 
		\caption{\fontsize{9.8}{11.8}\selectfont Kernel regression with different input and output dimensions, targeting higher-order parity functions. (a) Kernel regression to the fourth-order parity function \( f(x) = x_1 x_2 x_3 x_4 \) with binary inputs $(x_1, x_2, ..., x_M)$ of dimensions $M=9, 12, 14$. As the number of output speckle grains increases, the accuracy of kernel prediction increases, eventually reaching 100\%. A larger input dimension requires more speckle grains to perform the target function. (b) Kernel regression to the fifth-order parity function with input dimensions $M=9, 14$. The kernel accuracy for $M=14$ reaches 91.8\% with 4330 speckle grains (features), but for $M=9$ the accuracy stays around 50\% as the number of output features (speckle grains) increases.
		}
		\label{fig_kernel_by_grains}
	\end{figure}
    
	Figure~\ref{fig_kernel_by_grains}a shows the accuracy of kernel regression to the fourth-order parity function for varying input and output dimensions. For a fixed dimension $M$ of the input pattern on the DMD, the accuracy increases with the output speckle number and eventually reaches 100\%. For higher $M$, a larger number of output speckles is needed to reach 100\% accuracy. Although a smaller $M$ requires a lower output dimension to approximate the fourth-order parity function, it cannot approximate the fifth-order parity function accurately, no matter how large the output dimension would be. This is illustrated in Fig.~\ref{fig_kernel_by_grains}b, where the accuracy stays around 50\% for $M=9$. In contrast, the input dimension of $M=14$ can approximate the fifth-order parity function with 91.8\% accuracy when the number of output speckle grains is 4330. The accuracy will further increase with the output dimension. This result confirms that a larger input dimension can approximate more complex functions. Finally we also try to approximate the sixth-order parity function with $M = 14, 16$, but the accuracy remains low when the number of output speckle grains is increased to $\sim$4000.  
    
    
    We note that a linear scattering system with input encoded in the incident field pattern only has second-order nonlinearity when the output intensity pattern is measured. Hence, it cannot approximate parity functions of order exceeding 2. Our optical kernel based on the multiple-scattering cavity contains up to the fifth-order nonlinearity, and potentially even higher orders with a further increase of the number of scattering by the modulated scattering structure. 
	
	\section{Discussion and Conclusion} \label{conclusion}

    We have presented an optical kernel machine with programmable nonlinearity. It is implemented in a linear multiple-scattering cavity with spatial modulation of the scattering structure. By varying the number of times light is scattered by the modulated structure, the nonlinear strength is tuned continuously at constant low power. The kernel sensitivity and information capacity increase with the nonlinearity. This allows us to optimize the kernel nonlinearity to best approximate the parity functions of varying order for binary inputs. For sufficient input and output dimensions, our kernel can approximate the fifth-order parity function that is highly variable and nonlinear. 
    These results not only show the capability of our kernel in approximating complex nonlinear functions, but also reveal the power of adapting the kernel nonlinearity for diverse learning tasks.  

    Our kernel nonlinearity can be further enhanced by optimizing the cavity design to increase the photon lifetime or the DMD modulation area. This will allow the kernel to approximate more complex functions of higher-order nonlinearity, e.g., parity functions of sixth-order and higher. In the current experiment, the light input to the scattering cavity is monochromatic. If it is replaced by a broadband light or optical pulses, the feature space of the kernel can be further expanded by spectral or temporal multiplexing\cite{dong2025optical}. This will increase the expressive power of the kernel. Finally, real-time in situ optimization of structural nonlinearity can be incorporated into the training process. This will enable a low-power, programmable optical kernel with target-adaptive nonlinear response.

	
	
	\bmsection*{Acknowledgments}
	We acknowledge Logan Wright and Florent Krzakala for stimulating discussion. This work is supported partly by the Office of Naval Research under MURI Grant No.~N00014-24-1-2548.
	
	\bmsection*{Conflict of interest}
	
	The authors declare no conflict of interest.
	
	\bmsection*{Data Availability Statement}
	
	The data supporting the findings of this study are available from the corresponding author upon reasonable request.
	
	\printkeywords
	\bibliography{main}

\end{document}